\def\NN{N}
\def\aut#1{#1}
\def\aut#1{#1}
\def\Tr{{\rm Tr}}
\def\lfrac#1#2{{#1/#2}}
\def\xib#1{\mbox{\boldmath $\xi$}}      %

\documentstyle[pra,epsf,aps]{revtex}
\begin{document}
\sloppy
\title{Spiky Phases of Smooth Membranes.\\
Implications for Smooth Strings
}
\author{H.\ Kleinert\thanks{E-mail kleinert@physik.fu-berlin.de,~
URL http://www.physik.fu-berlin.de/\~{}kleinert
}
                   \\
         Freie Universit\"at Berlin\\
          Institut f\"ur Theoretische Physik\\
          Arnimallee14, D-14195 Berlin
     }
\maketitle
\begin{abstract}
We point out a possible mechanism
by which smooth surfaces
can become spiky
as the
constant of curvature stiffness
 $\kappa$  falls below certain
critical values. This happens
either in a single first-order
transition, or in a sequence of two Kosterlitz-Thouless-like
transitions. There may also be additional phases in which
the spikes
form a hexagonal solid-like array or a
disordered liquid-like
structure.
Our discussion suggests that there exist smooth strings between quarks.
\end{abstract}
{}~\\
\section{Introduction}
We would like to
point out
a possible rather universal mechanism
by which spiky superstructures can form on a membrane.
These can
undergo several interesting phase transitions,
whose nature we predict.

\begin{figure}[tbhp]
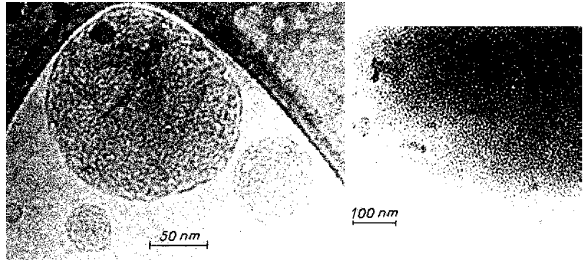

\vspace{3cm}
{}~~~~~~~~\input rough.tps
\caption[]{Rough Surface Structure observed by B. Kloesgen and W. Helfrich
 \cite{klos}
in dioleoylphosphatidylcholine (DOPC) bilayers.}
\label{eklos}\end{figure}

The existence of superstructures on membranes has been investigated
by many authors,
in particular wormhole
  \cite{sup1} and egg-carton shapes
  \cite{egg,fish}.
There are also
experimental  indications \cite{klos}
that such
structures exist (see Fig. \ref{eklos}).

\begin{figure}[tbhp]
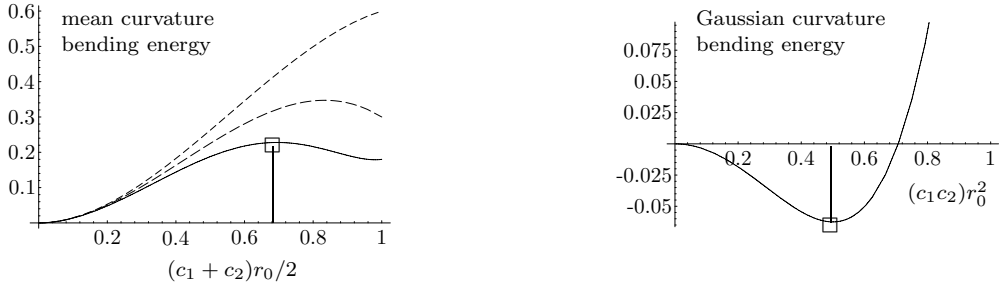

\vspace{2cm}
\input beh.tps
\caption[]{Possible behaviors of bending energy as a function
of mean curvature $(c_1+c_2)/2$
measured in units of the inverse molecular size $r_0$.}
\label{beh}\end{figure}

Consider closed amphiphilic
vesicles dispersed
in water, forming in general smooth tensionless surfaces \cite{5}
whose effective energy
is governed by curvature stiffness \cite{1}:
\begin{equation}
E = \frac{1}{2}\int d^2 x \sqrt{g}
  \left[   \kappa_0 (c_1 + c_2 )^2 + \bar \kappa _0
   c_1 c_2\right] ,
\label{ener}\end{equation}
 where $ \kappa_0$ is of the order
of eV. Here  $\lfrac{( c_1 + c_2)}{2 }$ and $c_1 c_2$
denote the mean and the Gaussian curvature, respectively,
$x=(x^1,x^2)$ are arbitrary parameters of the surface,
and $g_{ij}(x)$ is the intrinsic metric.
This energy is certainly only an approximation,
valid for small
curvatures.
If the curvatures increase,
there will be deviations from this simple quadratic behavior.
Moreover, since membranes are composed of rod-like molecules
of a certain length $r_0$,
there exists a natural maximal curvature $c_1=c_2=1/r_0$
beyond which a membrane cannot be bent without destroying the
microscopic structure.
Between zero and this maximal value,
the increase in energy
will slow down. This property is caused
by various important
contributions to the bending stiffness \cite{fish,elm}.
As a function of the mean curvature,
the bending energy of many  membranes
may have a maximum at a certain value $c_{\rm m}$
of $(c_1+c_2)/2$,
as
sketched in Fig.~\ref{beh}.
Such a maximum supplies the system with a second characteristic
length $r_{\rm m}=1/c_{\rm m}$ at which interesting new phenomena
should be observed.
A further length scale arises from higher powers of the Gaussian curvature
in the energy density.
Assuming, for example, the presence
of terms
$ \kappa_2(c_1c_2)^2+ \kappa_4(c_1c_2)^4$,
with coefficients
$ \kappa_2\approx- $eV\,\AA$^2$, $ \kappa_4\approx $eV\,(30\AA)$^8$,
Monte-Carlo simulations \cite{goe} have shown
the existence of a periodic egg carton superstructure
with a period of the order of $r_{\rm m}=60$\AA.

Membranes with an additional length scale
of either type
will be the objects of study in this note.
Certainly, the energy
of a single spike will be quite large (maybe of the order of $10k_BT$)
so that
their Boltzmann factor is quite small.
We shall see, however, that configuration entropy can
compensate this large energy at high enough temperatures.

\section{Reminder of Fluctuation Properties of Membranes}
{}~\\~\\[-1.5em]{\bf 2.}~
Before we come to our actual theory, let us
first recall
some well-known properties
of
membrane
 fluctuations
 governed only by the energy (\ref{ener}),
with a partition function (in natural units where the Boltzmann constant
is equal to 1)
\begin{equation}
Z= \sum_{\rm conf}e^{-E/T},
\label{@}\end{equation}
where $\sum_{\rm conf}$
denotes a sum
over all membrane configurations.
As a consequence
 of the
anharmonic nature of the energy (\ref{ener}), when expressed in terms of the
positions of the membrane molecules,
thermal fluctuations
make
the first coupling constant, the {\em extrinsic curvature stiffness\/},
  {\em soften\/} with temperature as \cite{2,2a}
\begin{equation}
    \kappa  =  \kappa _0 -3 (T / 8  \pi ) \log \left(
   r_{\rm IR}^2/r_{\rm UV}^2\right),
\label{1}\end{equation}
%
Here $ r_{\rm UV}$ is a short-distance (UV) cutoff equal to the
size
of the  molecules $r_0$, whereas $r_{\rm IR} $ is a long-distance (IR) cutoff
set by
the size of a vesicle.
The small parameter in the perturbation expansion leading to (\ref{1})
is the
inverse stiffness $1/ \kappa $, the {\em flexibility\/}.
The result (\ref{1})
is obtained from an infinite  bubble sum of diagrams,
which is equivalent to a Hartree-Fock-Bogoljubov
approximation \cite{2a}.
That approximation gives
 usually a good idea also for large
flexibilities, i.e. small $ \kappa $.
If we trust Eq.~(\ref{1}) in this regime,
we conclude
that the
extrinsic curvature vanishes
for large vesicles, whose size exceed
the so-called {\em  persistence length\/} \cite{4}
\begin{equation}
  \xi = r_{\rm UV}\,  e^{4 \pi  \kappa _0/3T}.
\label{3}\end{equation}
On the basis of this, we may expect vesicles of a size much larger than $\xi$
to
look crumpled.
This has indeed been confirmed in computer simulations \cite{cru}.

For the
 second coupling constant $\bar  \kappa_0$ in (\ref{ener}),
 the {\em  Gaussian curvature stiffness\/},
this effect is absent since this constant
 {\em hardens\/} as follows \cite{3}
\begin{equation}
  \bar  \kappa  = \bar  \kappa _0 + (10/3) (T/8 \pi )
\log \left(
   r_{\rm IR}^2/r_{\rm UV}^2\right) .
\label{2}\end{equation}
%

An important question
is whether
the persistence length (\ref{3})
reflects a true property of the theory,
or is merely a consequence of the one-loop approximation
which will not survive
higher-loop corrections.
In an attempt to answer this question,
consider
a single surface fluctuating around
an infinitely large planar configuration
in $d$ dimensions with periodic boundary conditions.
This surface may
be
 described by
a vector field $X^\mu( x)$ $(d=1,\dots,d)$,
giving rise to the
intrinsic metric
$g_{ij}(x)=\partial _i X^\mu(x)\partial _j  X_\mu(x)$
where
$(i,j=1,2)$. The contraction of the spatial vector indices $\mu$
is performed via the euclidean unit matrix $ G_{\mu \nu}= \delta_{\mu \nu}$.
In terms of $X^\mu(x)$,
the energy (\ref{ener})
can be written more explicitly as
\begin{equation}
 E =\frac{1}{2}\int d^2 x  \sqrt{g(x)}  \left[ \kappa _0    D^2
     X^\mu  (x) D^2 X_\mu  (x)+
 \bar\kappa_0
 \left(
D^2     X^\mu
D^2     X_\mu \!-\!D_  \nu D _ \lambda  X^\mu  D^  \nu D ^ \lambda
X_\mu\right)\right] ,
\label{oren}\end{equation}
where $D_\mu$ is the covariant derivative formed with the
help of the Christoffel symbol $ \Gamma_{ij}{}^k=g^{kl}(
\partial _ig_{jl}+
\partial _jg_{il}
-\partial _lg_{ij})/2$.
The
Gaussian curvature energy
 can be ignored,
since
it is
a constant
depending only on the genus of the surface.

We shall choose a special
parametrization
due to Gauss, in which
$X^1(x)=x^1,~X^2(x)=x^2$,
and $X^{a+2}(x)=u^a(x)$ with $(a=1,\dots,d-2)$,
so that
$g_{ij}(x)= \delta_{ij}+\partial _i u_a(x)\partial _j u^a(x)$,
with the contraction over indices $a$ being performed
via
the
$(d-2)\times (d-2)$-dimensional submatrix $ \delta_{ab}$ of $G_{\mu \nu}$.
Then we
rewrite
the energy (\ref{oren})
in yet another form
\cite{thd}
\begin{equation}
 E' = \frac{ \kappa _0}{2} \int d^2 x  \sqrt{h}
   \left[ D^2 X^a  D^2 X_a  +  \,\lambda ^{i j}  ( \partial_i
    X^a \partial_j X_a +  \delta_{ij}-h_{ij} )\right],
\label{7}\end{equation}
where
 $h_{ij} (x)$ and
$ \lambda _{ij} (x)$
are
two auxiliary
symmetric fields.
     The partition function of  the surface is
then given by the functional integral
\begin{equation}
  Z = \int {\cal D} u {\cal D} \lambda _{ij}   {\cal D} h_{ij}
	 e^{-E'/T}.
\label{7}\end{equation}
  The integral
 over $ \lambda _{ij}(x)$ ensures that the auxiliary field
$h_{ij} (x)$ coincides with the induced metric $g_{ij} (x)$,
and thus the
equality of
 $Z$ with the partition function
$ Z = \int {\cal D} X
	 e^{-E/T}$ for the original energy (\ref{oren}) [or (\ref{ener})],
if the parametrization is fixed likewise.
In the sequel we shall measure the stiffness constants in units
of temperature so that we can set $T=1$ everywhere.

  The advantage of the functional integral (\ref{7}) is that the
 surface positions $u^a (x)$ can be integrated out,
leaving a purely intrinsic effective energy
\begin{equation}
 E^{\rm eff} = \frac{d-2}{2}\left\{   {\rm Tr} \log \left[(D^2)^2 - D_i
\lambda ^{i j}
	D_j \right]  - \frac{ \kappa _0}{d-2} \int d^2 x  \sqrt{g}
       \lambda ^{ij} h_{ij}\right\}  .
\label{@Eeff}\end{equation}
  In the limit $d \rightarrow \infty$,  the fluctuations of the fields
 $ \lambda ^{ij} (x), h_{ij} (x) $ are frozen, and
the energy is given by the saddle point approximation.
For symmetry reasons, we may
assume the saddle point to have constant diagonal
fields
$h_{ij} =  \rho \, \delta _{ij},$
  $\lambda ^{ij} =  \lambda  h^{ij}
= (\lambda/ \rho) \delta_{ij}$,
 so that the energy
reads
\begin{equation}
E  = \frac{d-2}{2} \Delta x^1  \Delta x^2\, \rho
  \left\{
 \left[
\int \frac{d^2k}{(2\pi)^2} \log
    (k^4 +  \lambda k^2)
-2 \frac{\kappa _0 }{d-2} \lambda\right]
+ 2 \frac{\kappa _0 }{d-2} \frac{ \lambda}\rho
\right\} ,
\label{@en0}\end{equation}
 to be extremized in $ \rho $ and $ \lambda $.
Here $\Delta x^1  \Delta x^2$ is the base area of the surface.
Performing the integral with a momentum space cutoff $ \Lambda$ of the order of
the inverse
molecular size $r_{\rm UV}$
we find
\begin{eqnarray}
  \int_{\vert { k}\vert < \Lambda }
        \frac{d^2k}{(2\pi )^2} \log \left( { k}^2 + \lambda \right)
         & = & \frac{1}{4\pi } \left[ \left( \Lambda ^2+\lambda \right)
               \log \left( \Lambda ^2  + \lambda \right) - \Lambda ^2
               - \lambda \log \lambda \right] \nonumber \\
          & = & \frac{1}{4\pi } \left[ \Lambda ^2
                \left( \log\Lambda ^2 -1\right) + \left( \log \Lambda ^2
                +1 \right) \lambda - \lambda  \log \lambda \right]+{\cal O}(1/
\Lambda^2),
\label{si9.168}\end{eqnarray}
and the brackets in  (\ref{@en0}) become
\begin{eqnarray}
   \frac{1}{4\pi } \left[
   2       \Lambda ^2 \left( \log \Lambda ^2 - 1\right)
          + \left( \log \Lambda ^2 + 1 \right) \lambda
          - \lambda  \log \lambda \right] -
         2 \frac{\kappa _0 }{d-2}\lambda . \label{si9.169b}
\label{si9.169b}\end{eqnarray}
 The first constant term
can be absorbed into
 the measure of the functional integral.
The logarithmic divergence multiplying $ \lambda$
may be removed by introducing a
renormalized stiffness
\begin{equation}
 \kappa= \kappa_0-\frac{d-2}{8\pi}\left(\log \frac{\Lambda^2}{\mu^2}+1\right)
\label{@}\end{equation}
where $\mu$ is some mass scale, on which $ \kappa$ depends [$\kappa=
\kappa(\mu)$].
For $\mu$
equal to the inverse of the molecular size $r_{\rm UV}$,
we have  $ \kappa(r^{-1}_{\rm UV})= \kappa_0$.
With the help of $ \kappa$,
the bracket in (\ref{@en0}) becomes
\begin{eqnarray}
   -\frac{1}{4\pi }           \lambda
  \log\frac{\Lambda^2}{\mu^2}
 -
         2 \frac{\kappa  }{d-2}\lambda . \label{si9.169bx}
\end{eqnarray}
Introducing further a $\mu$-independent mass scale (the
so-called {\em dimensionally transmuted coupling constant\/})
\begin{equation}
\bar\lambda=\mu^2
e^{-[2/(d-2)]4\pi\kappa(\mu)+1
},
\label{@}\end{equation}
expression (\ref{si9.169b})
takes a $\mu$- and cutoff-independent form
\begin{eqnarray}
   -\frac{1}{4\pi }           \lambda
  \log\frac{ \lambda}{\bar \lambda}
  , \label{si9.169bxy}
\end{eqnarray}
and the energy (\ref{@en0})
can be rewritten as
\begin{equation}
E  = \frac{d-2}{2} \Delta x^1  \Delta x^2\, \rho        \left[
 f_0-\frac{ \lambda}{4\pi}
+ \kappa\frac{ \lambda}{ \rho}
\right] ,~~~~f_0\equiv -\frac{ \lambda}{4\pi}\left( \log\frac{\lambda}{\bar
\lambda}-1\right) .
\label{@tobe}\end{equation}
A multiplicative renormalization
factor $Z_ \kappa=\kappa_0/ \kappa$
 has been absorbed into $ \lambda$ and $\bar  \lambda$,
so that all quantities are now finite
for $ \Lambda\rightarrow \infty $.
Extremizing (\ref{@tobe})
in $ \rho$ yields
$f_0= \lambda/4\pi$ and thus $ \lambda= \bar \lambda$,
where $f_0- \lambda/4\pi=0$.
Extremizing (\ref{@tobe})
 in $ \lambda$ yields  $\kappa/ \rho=1/4\pi$,
so that the extremal energy (\ref{@tobe})
is
\begin{equation}
E^{\rm ext}= \frac{d-2}{2} \Delta x^1  \Delta x^2\,\frac{\bar  \lambda}{4\pi}.
\label{@}\end{equation}
{}From the trace of the logarithm in
(\ref{@Eeff})
we see that $\bar  \lambda$
sets
a mass scale
for the
correlation function
$\langle D_iu^a D_ju^b\rangle $
which falls off like $e^{-|x| \sqrt{\bar  \lambda}}$ for large $|x|$,
showing that $1/ \sqrt{ \bar  \lambda}$
plays the role of the persistence length
(\ref{3}).
Thus the  $d\rightarrow \infty$ model
possesses precisely the properties
which we derived for a real membrane in
three dimensions
by a one-loop approximation.

The question of corrections to the
one-loop approximation is therefore
equivalent to the problem
of lowering $d$
down to the physical dimension $3$.
Since we are unable to treat the model
for finite $d$,
let us gain insight into its possible properties
by comparing it
to
an analogous
very similar
model, which
is also exactly solvable for a parameter
$\NN\rightarrow \infty$,
which plays the same role as $d-2$ in the above discussion.
This is the O($\NN$) nonlinear
$ \sigma$-model
for which
the case $\NN=2$ is well-known
to have quite different properties
from those derived from the $\NN\rightarrow \infty $ -limit.
This knowledge
will shed some light upon the behavior of surfaces
for small $d-2$.

\section{Relevant Properties of O($N$)-Symmetric
Nonlinear $ \sigma $-Models}
The O($\NN$)-symmetric  nonlinear
$ \sigma$-model consists of a fluctuating field of
unit vectors
 with $\NN$ components
\begin{eqnarray}
   n_a (x)~~(a = 1, \dots , \NN),~~~~~ n_a^2(x) = 1,
\label{9.67}\end{eqnarray}
in a two-dimensional $x$-space,
with a partition function given by the functional integral
\begin{eqnarray}
  Z_{\sigma} = \prod _{{ x}} \left[ \int \frac{d^{\NN-1}n_a}{S_\NN}\right]
               \exp\left[ -\frac{ \kappa_0}{2} \int d^2 x (\partial
n^a)^2\right] ,
\label{si9.138}\end{eqnarray}
where $S_\NN\equiv 2 \pi  ^{\NN/2}/\Gamma (\NN/2)$
 is the surface of a sphere in $\NN$ dimensions
  covered by the directional integral $d^{\NN-1} n^a$.
This ensures a unit integral $ \int
      d^{\NN-1} n^a/ S_\NN = 1$.
Just as the surface model
for $d\rightarrow \infty$,
this model
is exactly solvable
in the limit $\NN \rightarrow {\infty}$, where it is
 referred to as the {{\em spherical model}},
 first solved by Berlin and Kac in 1952 \cite{Kac}. As in the energy
(\ref{7}), we introduce
an auxiliary
field $ \lambda(x)$
and rewrite (\ref{si9.138})
as
\begin{eqnarray}
  \!\!\!\!\!\! Z_{\sigma}= \int {\cal D}^\NN n_a\int_{-i\infty }^{i\infty }
{\cal D}  \lambda
        \exp \left\{ - \frac{ \kappa_0}{2} \int d^2x
              \left[ \left( { \partial} n_a\right) ^2
               +\NN \,\lambda  \left( n_a^2 - 1\right) \right] \right\},
\label{si9.144}\end{eqnarray}
where the $ \lambda$-integrations
run from $-i\infty $ to $i\infty $.
Now the $n_a$-integrals
are Gaussian and can be done, leading to
\begin{eqnarray}
 \!\!\!\!\!\!\!  Z_{\sigma} =
\int_{-i\infty }^{i\infty }    {\cal D}  \lambda
    \exp \left\{ - \frac{\NN}{2} \left[ \Tr \log
                \left( -\partial ^2 + \lambda\right)  +
       \kappa_0 \int d^2 x \,\lambda ({ x}) \right]\right\} .
\label{si9.145}\end{eqnarray}
In the limit $\NN \rightarrow {\infty} $, the
$ \lambda$-fluctuations are frozen at the saddle point, which
 lies at a constant $ \lambda(x)\equiv  \lambda$, and has
an
energy
\begin{equation}
E_{ \sigma}= \frac{\NN}2\Delta x^1  \Delta x^2  \left[
\int \frac{d^2k}{(2\pi)^2} \log
    (k^2 +  \lambda )
- \frac{\kappa _0}\NN  \lambda\right].
\label{@}\end{equation}
Regularizing
the integral as before,
this becomes
\begin{equation}
E  = \frac{\NN}{2} \Delta x^1  \Delta x^2\, \rho
 f_0
 ,~~~~f_0\equiv -\frac{ \lambda}{4\pi}\left( \log\frac{\lambda}{\bar
\lambda}-1\right) ,
\label{@tobe1}\end{equation}
where $ \bar\lambda=\mu^2
\exp\{-4\pi\kappa(\mu)/\NN\} $
is the dimensionally transmuted coupling constant
of the $ \sigma$-model.
It gives a nonzero length scale $1/  \sqrt{ \bar\lambda}$
to the fluctuations of the vector field $n_a$
which have a correlation function $\propto e^{-|x| \sqrt{\bar\lambda} }$.

The similarity of the two models
is quite obvious.
The advantage of the latter
is that
it is much better understood.
Since the work of Kosterlitz and Thouless in 1973 \cite{seeKT}
it is known that $\NN\rightarrow \infty$ properties
are found in the model only down to $\NN=3$ (classical Heisenberg model).
For $\NN=2$,
the model
possesses
at large $ \kappa_0\ge  \kappa_0^c=2/\pi$ an
extra
 phase in which the correlation functions
of the
 vector field $n_a$
have a long range, falling off
algebraically like
 $1/|x|^{\rm const}$
rather than exponentially.
This is most easily seen by parametrizing the two-component vector
field $N_a$
in terms of an azimuthal angle $\theta$ as $(N_a)=(\cos\theta,\sin \theta)$,
and rewriting
(\ref{si9.138})
as
a functional integral
\begin{eqnarray}
  \tilde Z_{\sigma} =\sum _{L}  \prod _{{ x}} \left[ \int \frac{d\theta({
x})}{2\pi}\right]
               \exp\left\{ -\frac{ \kappa_0}{2} \int d^2 x [\partial
_i\theta(x)
-2\pi \delta_i(x;L)
]^2\right\} ,
\label{si9.138b}\end{eqnarray}
where $ \delta_i(x;L)\equiv  \epsilon_{ij}\int d\xi_j
\delta^{(2)}(x-\bar x(\xi)) $
is the $ \delta$-function on a line $L$ described by $\bar x(\xi)$,
pointing orthogonal to the line elements. The sum
$\sum _{L}$ runs over a grand-canonical ensemble
of lines $L$.
The sum is necessary
to preserve
the cyclic invariance of the energy in (\ref{si9.138}) under
cyclic replacements $\theta(x)\rightarrow \theta(x)+2\pi  \delta(x;S)$,
where  $ \delta(x;S)\equiv  \int d^2\xi
\delta^{(2)}(x-\bar x(\xi))) $
is the
 $ \delta$-function on the surface $S$ described by $\bar x(\xi)$
\cite{camlec}.
The energy in (\ref{si9.138b})
is a direct consequence of the
identity $ \partial _i e^{i\theta(x)}=e^{i\theta(x)}
[\partial _i\theta(x)
-2\pi \delta_i(x;L)]$.

We may now introduce an auxiliary field
$b_i(x)$ and rewrite $\tilde Z_{\sigma}$ as
\begin{eqnarray}
  \tilde Z_{\sigma} =\sum _{L}  \prod _{{ x}} \left[ \int \frac{d\theta({
x})}{2\pi}\int db_i\right]
               \exp\left\{{ \kappa_0}
 \int d^2 x\,\left(
\frac{b_i^2(x)}{2}-i \,b_i(x) [\partial _i\theta(x) -2\pi \delta_i(x;L)]\right)
\right\} .
\label{si9.138bc}\end{eqnarray}
The integrals over $\theta(x)$
can now be performed,
which gives the condition that the field $b_i$ must be divergenceless
\cite{camlec}.
This may be enforced by setting
$b_i= \epsilon_{ij}\partial _j u$,
and the partition function
(\ref{si9.138bc})
becomes
\begin{eqnarray}
  \tilde Z_{\sigma} =\sum _{\{x_i,x_f\}}  \prod _{{ x}} \left[ \int{du({x})}
\right]
               \exp\left\{ \kappa_0 \int d^2 x\,\left[
\frac{1}{2}[\partial _iu(x)]^2
-2\pi iu(x)n(x)
\right]
\right\} ,
\label{si9.138bd}
\end{eqnarray}
where
\begin{equation}
n(x)=  \epsilon_{ij}\partial _i  \delta_j(x;L) = \delta^{(2)}(x-x_i)-
\delta^{(2)}(x-x_f)
\label{@}
\end{equation}
is the density of initial and final
end points of the line $L$, which lie at $x_i$ and $x_f$, respectively,
with positive and negative signs.
These are the positions of vortices and antivortices
in the original
field configurations
$\theta(x)$.
By integrating out
the $u$-field in (\ref{si9.138bd}), we obtain
\begin{eqnarray}
  \tilde Z_{\sigma} =\sum _{\{x_i,x_f\}}
               \exp\left[ -\frac{\kappa_0}{2}4\pi^2
\int d^2 x\,
\int d^2 x'\,
n(x)G(x-x') n(x')
\right],
\label{si9.138bd1}\end{eqnarray}
where
\begin{equation}
G(x-x')=\int \frac{d^2k}{(2\pi)^2}\frac{e^{i{ k}{ x}}}{k^2}=
-\frac{1}{2\pi}
\log\frac{|x-x'|}{r_0}+\log \frac{r_{\rm IR}}{r_0^2}~~~~
\label{@}\end{equation}
is the correlation function of the field $u(x)$
containing some finite length scale $r_0$
and
an infrared cutoff $r_{\rm IR}$ which
goes to infinity.
However, this infinity drops out in
the partition function
(\ref{si9.138bd1}) since
$\int d^2x\, n(x)=0$
due to the equal number of
initial and final endpoints
$x_i,x_f$ in the
sum over all initial and final points
$\Sigma _{x_i,x_f}$
of  the lines $L$
 in the grand-canonical ensemble.

\begin{figure}[tbhp]
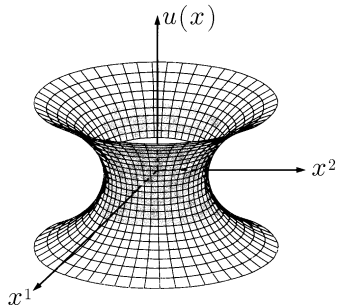

\vspace{-1cm}
{}~~\input fig35.tps
\caption[Wormhole solution of differential equation $\partial
^2u(x)=0$]{Wormhole solution
of differential equation  $\partial ^2u(x)=0$.}
\label{wh}\end{figure}

\begin{figure}[tbhp]
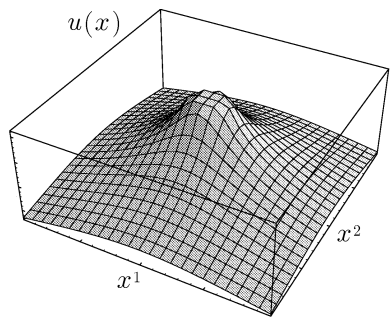

\input mon.tps
\caption[]{Spike-like solution of  differential  equation
$\partial ^2 u(x)=0$, obtained from the lower branch
of the ``wormhole" solution
in Fig.~\ref{wh} after capping it at the molecular scale $r_{\rm UV}$.
}
\label{mono}\end{figure}

\begin{figure}[tbh]
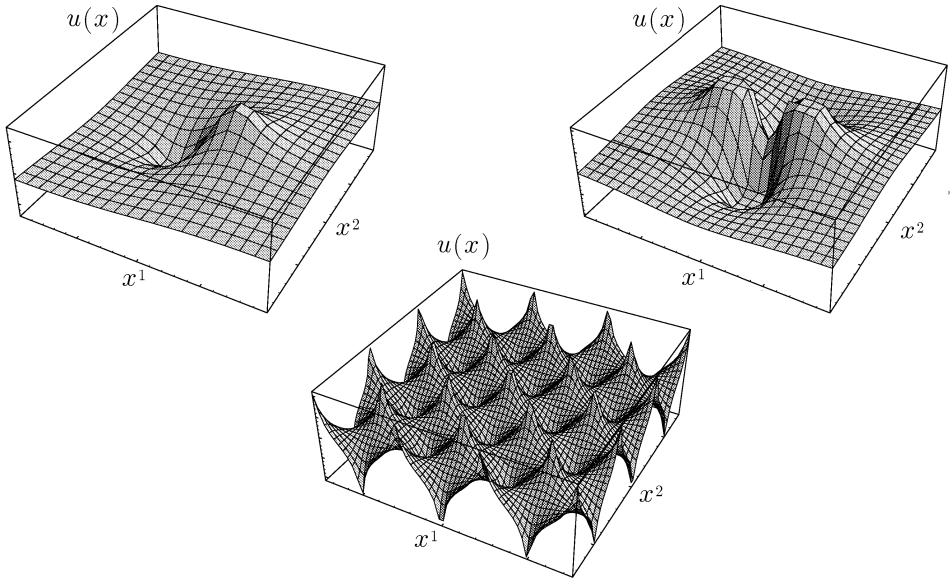

\input dipqu.tps
\vspace{1.8cm}
 \caption[Pair of spikes, quadrupole-,
and crystal-like multipole solutions, obtained from a superposition of
spike and antispike
solutions
of the differential equation $\partial ^2u(x)=0$ in Fig.~\protect\ref{mono}]
{Pair of spikes, quadrupole-,
and crystal-like multipole solutions, obtained
from a superposition of
spike and antispike
solutions
of the differential equation $\partial ^2u(x)=0$ in Fig.~\ref{mono}.
}
\label{dip}\end{figure}

Let us derive the partition function
(\ref{si9.138bd1}) in a slightly different, more phenomenological
way
which will be useful to understand the smooth-membrane problem.
Suppose we are given the partition function
(\ref{si9.138bd}) without vortices, where the energy is simply
\begin{equation}
\tilde E{}^{\rm lin}_ \sigma=\frac{ \kappa_0}{2}\int d^2x (\partial_i  u)^2.
\label{@field0}\end{equation}
The equation of motion
\begin{equation}
\partial ^2 u(x)=0
\label{@equm0}\end{equation}
has two types of rotationally invariant solutions. The trivial $u(x)\equiv 0$,
and the solution
\begin{equation}
u_{\rm wh}(x)= \mp A\frac{1}{2\pi}\log\frac{r}{r_0},~~~~r>r_0,~~~~r\equiv |x|,
\label{@worm}\end{equation}
which satisfies $-\partial ^2u_{\rm wh}
 =0$. The amplitude $A$
is arbitrary.
If we interpret $u(x)$ as a vertical displacement of a flat surface
in three dimensions, this looks like a ``wormhole" in
a curved planar space, as
pictured in Fig.~\ref{wh}.
This solution exists for any $r_0$.
Let us now imagine the surface
to consist of molecules of size $r_{\rm UV}$.
Then the smallest possible $r_0$
is $r_{\rm UV}$. For this value, however,
the throat of the wormhole
can be closed by a small cap from above or below,
thus
becoming a spike-like surface shown in Fig.~\ref{mono}
to be denoted by $u_{\rm sp}(x)$, or an antispike-like
surface pointing downwards.
The linearized energy
(\ref{@equm0}) of these surfaces is large so that
only a few spikes and antispikes are present in thermal
equilibrium.
If a single  spike-antispike pair
with opposite orientation
centered at different places $x$ and $y$ is
inserted into the field
 energy
(\ref{@field0}), we obtain
precisely the interaction energy
of a vortex pair in the partition function (\ref{si9.138bd1})
if
 we set $A=1$.
The size of $A$ is a consequence of the
vortex quantization in the original
energy in (\ref{si9.138b}).
A spike
and an antispike can combine to a dipole, and two of these
to a quadrupole, as shown in Fig.~\ref{dip}.
In principle, there
can also be a
crystal-like array of
spikes and antispikes, also
shown in Fig.~\ref{dip}. This requires the presence
off
additional
higher-gradient terms in the energy
(\ref{@equm0}) which would
modify the short-distance properties appropriately \cite{camlec}.

The partition function (\ref{si9.138bd1})
describes a neutral Coulomb gas.
This is known to have
a pair unbinding transition
which is of infinite order at
low
temperatures, the Kosterlitz-Thouless transition.
The transition point is easily calculated
if only very few charges are present.
At low temperatures,
a single pair has an average square distance
\begin{equation}
\langle r^2\rangle \propto \int_{r_0}^\infty  dr\,r \,e^{- \kappa_0\,2\pi
\log(r/r_0)}
r^2\propto\frac{1}{4- \kappa_0\,2\pi}.
\label{@}\end{equation}
As the stiffness constant
$ \kappa_0$ falls below
$ \kappa_0^c=2/\pi$,
the square distance diverges, and the electron gas enters a plasma phase
in which
the field $u(x)$ acquires a finite range,
the {\em Debye\/}
screening length $r_{\rm Debye}$,
which
is precisely the inverse of
 $ \sqrt{ \bar  \lambda}$ in Eq.~(\ref{@tobe1}).

 This result implies that
the average
value of $ \lambda(x)$
which is calculated
with
the functional integral
(\ref{si9.145})
as
\begin{eqnarray}
 \!\!\!\!\!\!\!  \langle \lambda(x)\rangle =Z^{-1}
\int_{-i\infty }^{i\infty }    {\cal D}  \lambda
 \,\lambda(x)  \,  \exp \left\{ - \frac{\NN}{2} \left[ \Tr \log
                \left( -\partial ^2 + \lambda\right)  +
       \kappa_0 \int d^2 x \,\lambda ({ x}) \right]\right\} ,
\label{si9.145x}\end{eqnarray}
and which has the value $ \langle \lambda(x)\rangle=\bar  \lambda\neq 0$
for small stiffness,
vanishes for $\NN=2$ and $ \kappa_0 > \kappa_0^c$.
The fluctuations of $ \lambda(x)$
become so strong
that
the nonzero saddle point at $ \lambda\equiv \bar  \lambda$
looses control over the functional integral.
The $ \lambda(x)$-field will acquire more and more zeros
until, at a small enough $ \NN$ and large enough stiffness,
these zeros proliferate making $\langle  \lambda(x)\rangle =0$.
Up to now, nobody has been able to derive
this result from
the functional integral (\ref{si9.145x}),
but the above Coulomb gas argument
proves that it must be true.

For a dilute gas of vortices,
i.e., for a high activation energy $E_{\rm v}$
and a small fugacity $z=e^{-E_{\rm v}/T}$
of a vortex,
the sum over all vortices and antivortices
in the partition function
(\ref{si9.138bd}) can be restricted to a single
vortex or antivortex
at each point, in which case the
partition function goes over into
\begin{eqnarray}
  \tilde Z_{\sigma} \approx \prod _{{ x}} \left[ \int{du({x})} \right]
               \exp\left\{- \kappa_0 \int d^2 x\,\left[
\frac{1}{2}[\partial _iu(x)]^2
-\frac{z}{ \kappa_0}\cos[2\pi \kappa_0 u(x)]
\right]
\right\} .
\label{si9.138bdx}
\end{eqnarray}
This is the partition function
of the
sine-Gordon model,
which is thus
equivalent to the O(2) nonlinear $ \sigma$-model.
For small stiffness $ \kappa_0$,
we may expand
$\cos[2\pi  \kappa_0u(x)]\approx1-4\pi^2 \kappa_0^2u^2(x)/2$
to see
that the fluctuations have a a finite range.
For large stiffness $ \kappa_0>2/\pi$, however,
the cosine oscillates so rapidly that it no longer
contributes to the functional integral.
In this regime, the
system has only
long-range fluctuations \cite{cole}.

\section{Consequences for Membranes}
What are the lessons of all this for membrane fluctuations?
Suppose we restrict the field fluctuations
in  (\ref{7})
to diagonal fields $h_{ij}(x)= \rho(x) \delta_{ij}$ and $ \lambda^{ij}=
 \lambda(x)h^{ij}(x)$
with the constant extremal $ \rho=4\pi  \kappa$.
Then we remain with a functional integral
over $ \lambda(x)$ with a Boltzmann factor
$e^{-E'{}^{\rm eff}/T}$ where $E'{}^{\rm eff}$
is the obvious  generalization of the effective energy (\ref{@Eeff})
to a functional of a space-dependent $ \lambda(x)$
which has precisely the same form as functional
the exponent in
(\ref{si9.145}).
But such a functional integral
has just been shown
to yield for a large enough stiffness $ \kappa$
a vanishing average $\langle  \lambda(x)\rangle $,
implying that the fluctuations of the
derivatives
$\partial _iu^a$
have an infinite range
characteristic for a smooth surface.
It is certainly conceivable that the neglected fluctuations, nondiagonal in $
\lambda^{ij}$
and arbitrary local in $h_{ij}(x)$
do not change this  result.

\begin{figure}[tbhp]
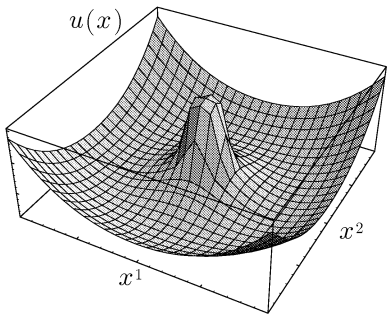

\input mon2n.tps
\caption[]{Spike-like solution of  equation
$\partial ^4 u(x)=0$, with singularity capped at the molecular scale $r_{\rm
UV}$.
}
\label{mon2}\end{figure}

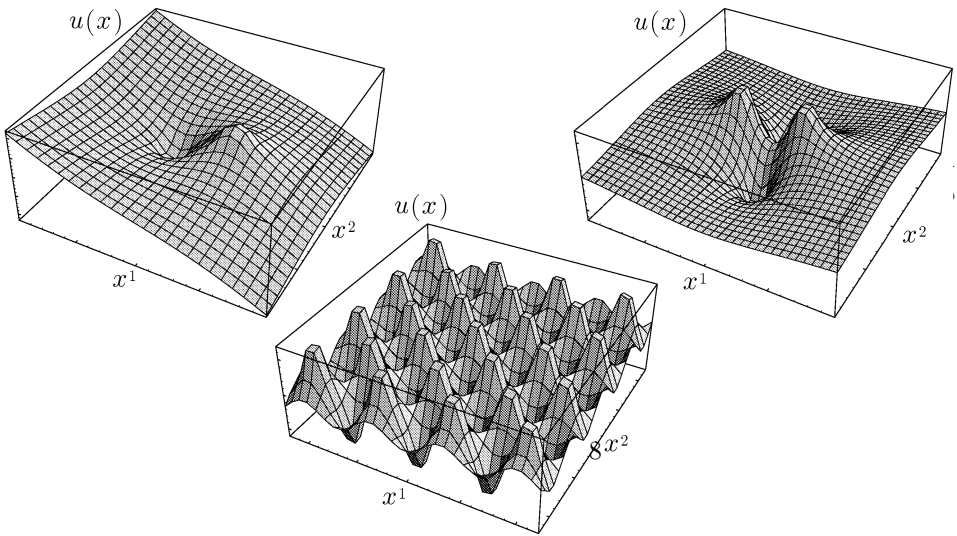
\begin{figure}[tbhp]
\input dipqu2.tps
\caption[Pair of spikes, quadrupole and multipole solutions]{Pair of spikes,
quadrupole, and multipole solutions
of differential equation $\partial ^4u(x)= 0$,
with singularities
capped at
the molecular scale $r_{\rm UV}$.
}
\label{dip2}\end{figure}

How can we see whether this is  true?
Consider a linearized version
of the energy (\ref{oren})
\begin{equation}
\tilde E{}^{\rm lin}=\frac{ \kappa_0}{2}\int d^2x [\partial ^2 u(x)]^2,
\label{@ensp}\end{equation}
and let us go through the same argument as
in the discussion of
the linearized energy (\ref{@field0}).
The equation of motion
\begin{equation}
\partial ^4 u(x)=0
\label{@}\end{equation}
has three types of solutions. First, there is the trivial one $u(x)\equiv 0$.
Second, there are
spike-like solutions $u_{\rm sp}(x)$
arising from capping a wormhole solution
(\ref{@worm}) of the previous system,
as shown in Fig.~\ref{wh}.
Third, and most importantly for us,
there
are spike-like solutions of the form
\begin{equation}
u_{\rm sp2}(x)=\pm A\frac{ \sqrt{3} }{2}\frac{1}{8\pi}\left[
r^2\log\frac{r}{r_0e}
-\frac{1}{2}\log\frac{r}{r_0e^{1/6}}\right],
{}~~~~r\equiv |x|,
\label{@spike2}\end{equation}
which
satisfy $-\partial ^2u_{\rm sp2}(x)
=u_{\rm sp}(x)$
and $\partial ^4u _{\rm sp2}(x) =\pm A \delta^{(2)}(x)$.
With a cap of molecular size they
have the form shown in Fig.~\ref{mon2}.
There are also dipole, quadrupole, and multipole solutions
displayed in Fig.~\ref{dip2}.
The distances of the spikes in these arrays are of the order of the length
scale
$r_{\rm m}=1/c_{\rm m}$ supplied by the extrema in Fig.~\ref{beh}.
The important point is now that
there must be a preferred amplitude $A$ of the spike-like solution
(\ref{@spike2}),
fixed by the interplay between
$r_{\rm m}$ and
the molecular
size $r_{\rm UV}$ (as in the Monte-Carlo calculation in Ref.~\cite{goe}).
If this is the case, we
can now insert a superposition of these solutions
into the energy
(\ref{@ensp}) and find a partition function
\begin{eqnarray}
  \tilde Z_{\rm mem} =\sum _{\{x_i,x_f\}}
               \exp\left[ - \frac{\kappa_0}{2}A^2
\int d^2 x\,
\int d^2 x'\,
n(x)G_4(x-x') n(x')
\right],
\label{si9.138bd1v}\end{eqnarray}
where
$n(x)$ is a sum of $ \delta$-functions
at the positions of the spikes (with a minus sign for antispikes), and
\begin{equation}
G_4(x-x')=\int \frac{d^2k}{(2\pi)^2}\frac{e^{i{ k}{ (x-x')}}}{k^4}=
u_{\rm sp2}(x-x')+ c_1(x-x')^2+c_2
\label{@}\end{equation}
is the correlation function of the field $u(x)$
containing two infinite constants
from the infrared divergence of the integral.
The infinity $c_2$ drops out for neutral spike-antispike systems
with $\int d^2x\,n(x)=0$,
the infinity $c_1$ drops out under the
condition of {\em dipole neutrality\/}
$\int d^2x\,x\,n(x)=0$.
The latter follows directly from
\begin{eqnarray}
\int d^2x\,\int d^2x'\,n(x)\,(x-x')^2n(x')
&=&2\int d^2x\,x^2\,n(x)\int d^2x'\,n(x')
-2\int d^2x\,x\,n(x)\int d^2x'\,x'\,n(x')\nonumber \\&=&
-2\int d^2x\,x\,n(x)\int d^2x'\,x'\,n(x')=0.
\label{@}\end{eqnarray}
Note that the simpler spike-like solutions $u_{\rm sp}(x)$
in Fig.~\ref{mono} do not have a long-range attraction between them.

The partition function
(\ref{si9.138bd1v})
is obviously equivalent to the
local one
\begin{eqnarray}
  \tilde Z_{\rm mem}  =\sum _{\{x_i,x_f\}}   \prod _{{ x}} \left[ \int{du({x})}
\right]
               \exp\left\{ \kappa_0 \int d^2 x\,\left[
\frac{1}{2}[\partial^2u(x)]^2
-A iu(x)n(x)
\right]
\right\} ,
\label{si9.138bdxx}
\end{eqnarray}
or, in analogy with
(\ref{si9.138bdx}), to
\begin{eqnarray}
  \tilde Z_{\rm mem} \approx \prod _{{ x}} \left[ \int{du({x})} \right]
               \exp\left\{- \kappa_0 \int d^2 x\,\left[
\frac{1}{2}
[\partial^2u(x)]^2
-\frac{z}{ \kappa_0}\cos[A \kappa_0 u(x)]
\right]
\right\} ,
\label{si9.138bdxx}
\end{eqnarray}
This system is the dual transform
of
a two-dimensional crystal with defects whose
phase properties have been thoroughly studied
\cite{GaugeF}.
As it stands, the partition function (\ref{si9.138bdxx})
has a single first-order
transition \cite{langh}
from a spiky to a smooth surface.
If the short-range properties of the model are slightly modified
at a
length scale $l$,
the first-order transition splits into
a sequence of two Kosterlitz-Thouless-like transitions,
in which the spikes first form a gas of dipole pairs,
and then a gas of quadrupoles \cite{twostep} (see
the figure on
p. 1303 of Ref.~\cite{GaugeF}).

Upon further modifying the short-range properties,
the quadrupoles
may combine to a liquid of
spikes and antispikes,
which may be
the structures
seen in the experiments in Fig.~\ref{eklos}.
The liquid
could also freeze to a solid.
Future experiments
should look for quadruplets and dipoles to confirm the
correctness of these ideas.

It is essential that nonlinearities in the curvature energy
fix an optimal size of the amplitude $A$ in
(\ref{si9.138bd1v}).
For a purely quadratic  surface energy
(\ref{ener})
such an optimal $A$ cannot exist
since the energy of a spike-like solution
(\ref{@spike2})
can be reduced continuously to zero by letting $A$ go to zero.

\section{Possible Consequences for String Theory}
Such nonlinearities
may also resolve an
an old riddle
in the string theory of
{\em permanent quark confinement\/}.
Quarks and antiquarks are
held together
at all distances
by
color-electric flux tubes formed
from nonabelian gauge fields of quantum chromodynamics.
In a euclidean spacetime with $d=4$, these tubes form fluctuating surfaces.
In string theory,
the
behavior of
such surfaces is investigated
in an idealized form,
considering them as
infinitely thin objects
possessing initially only surface tension
({\em Nambu-Goto strings\/}).
But soon it was found
that these surfaces would be unstable against the
formation of infinitely thin protrusions,
which do not possess any surface energy
but a large configurational entropy  ({\em plumber's nightmare\/}).
This instability would ruin quark confinement.
To avoid this,
a curvature stiffness energy
was
added to the surface tension \cite{polkl}.
The resulting
{\em stiff-string\/}
action was equal to
the energy
(\ref{7}).
However, this did not seem to lead to the desired stabilization
since,
in an infinite number of dimensions,
there exists
a finite persistence length
beyond which quark would no longer
be confined. It remained unclear whether this
phenomenon would persist
down to $d=4$.
The present discussion
gives rise to the hope
that the persistence length
can become infinite after all. Since $d=4$ in (\ref{7}) corresponds to $N=2$
in the nonlinear $ \sigma$-model (\ref{si9.145}), which
 possesses a smooth
phase with long-range
correlations,
the $ \lambda$-fluctuations
in the action (\ref{@Eeff})
may be violent enough
to produce so many zeros in $ \lambda(x)$
that $\langle  \lambda(x)\rangle =0$
and
the
persistence length in
 infinite,
implying permanent quark confinement.
But we also have learned
that this may occur only
with support
from higher nonlinear terms
in the curvature energy. A
color-electric flux tube certainly has such additional
terms, although
it differs from the membranes
discussed in this paper
in an important point:
Its curvature stiffness is negative,
as was recently shown by the author and Chevyakov \cite{klch}.

Papers by the author
in the list of references
can be read on the internet,
for instance (pubs1\#181) under
the www address
http://www.physik.fu-berlin.de/\~{}kleinert\/kleiner\_re1\#181.

{}~\\~\\
Acknowledgement\\
The author is grateful to
Prof. W. Helfrich and to
Drs. C. Diamantini, A. Pelster, and C. Trugenberger for
many
discussions.


\begin{thebibliography}{11}
\bibitem{sup1}
 M.M. Kozlov and W. Helfrich,
J. Phys. II (France) {\bf 4}, 1427 (1994),
Phys. Rev. E {\bf 51}, 3324 (1995);
B. Fourcade, Ling Miao, M. Rao, M. Wortis, and R.K.P. Zia,
Phys. Rev. E {\bf 49}, 5276 (1994);
U. Seifert, J. Shillcock, P. Nelson,
 Phys. Rev. Lett. {\bf 77}, 5237 (1996);
T. Charitat and B. Fourcade,
J. Phys. II (France) {\bf 7}, 15 (1997).

\bibitem{egg}
J.B. Fournier, Phys. Rev. Lett. {\bf 76}, 4436 (1996);
\bibitem{fish}
T. Fisher,
J. Phys. II (France) {\bf 2}, 337 (1992);
{\bf 3}, 1795 (1993).
\bibitem{elm}
M. Kiometzis and H. Kleinert,
      Phys.\ Lett.\ A {\bf 140}, 520 (1989) (pubs1\#200);
H. Kleinert,
    Phys.\ Lett.\ {\bf B211}, 151(1988) (pubs1\#177);
    Phys.\ Lett.\ A {\bf 136}, 253(1989) (pubs1\#181).

\bibitem{cru}
G. Gompper and D.M. Kroll, Europh. Lett. {\bf 19}, 581 (1992),
Science {\bf 255}, 968 (1992).


\bibitem{klos}
 B. Kl\"osgen and W. Helfrich,
Eur. Biophys. J. {\bf 22}, 329 (1993);
Biophys. J. {\bf 73}, 3016 (1997).
\bibitem{5}
 H.J.\ Deuling and W.\ Helfrich, J.\ Phys.\ France {\bf 37},
  1335, (1976);\\
W. Harbich,  H.J.\ Deuling and W.\ Helfrich, J.\ Phys.\ France {\bf 38},
  727, (1976).

%
\bibitem{1}
P.B. Canham, J.\ Theor. Biol. {\bf 26}, 61 (1970);\\
W.\ Helfrich, Z.\ Naturforsch.\ B {\bf 103}, 167 (1975).
\bibitem{goe}
R. Goetz and W. Helfrich, J. Phys. II France {\bf 6}, 215 (1996).
\bibitem{2}
 L.\ Peliti and S.\ Leibler, Phys.\ Rev.\ Lett.\ {\bf 54},
 1690 (1985).
\bibitem{2a}
H.\ Kleinert, Phys.\ Lett. A {\bf 114}, 263 (1985)
%
(pubs2\#128;
\bibitem{4} P.-G.\ de Gennes and C.\ Taupin, J.\ Phys.\ Chem.\
 {\bf 86}, 2294 (1982).
\bibitem{3}
 H.\ Kleinert, Phys.\ Lett.\ A {\bf 116}, 57 (1986)
(pubs2\#126;\\
See also:\\
 H.\ Kleinert and S. Ami,  Phys.\ Lett.\  A {\bf 120}, 207 (1987)
(pubs2\#157);\\
H. Kleinert,
    J. Stat.\ Phys.\ {\bf 56}, 227 (1989)
(pubs2\#175);\\
B.D. Simons and M.E. Cates,
J. Phys. II France {\bf 2}, 1439 (1992);                       \\
D.C. Morse and S.T. Milner, Europhys. Lett. {\bf 26}, 565 (1994).

\bibitem{thd}
A. Polyakov, Nucl. Phys. B {\bf 268}, 406 (1986).\\
H. Kleinert,
     Phys.\ Lett.\  B {\bf 174}, 335 (1986)
(pubs2\#149);
Phys. Rev. Lett. {\bf 58}, 1915 (1987)
(pubs2\#164);
Phys. Lett. B {\bf 189}, 187 (1987)
(pubs2\#162);
Phys.\ Rev.\ D {\bf 40}, 473 (1989)
(pubs2\#188).

%
\bibitem{Kac}
\aut{T.H. Berlin} and \aut{M.~Kac},
 Phys. Rev. {\em 86\/}, 82 (1952).
\bibitem{seeKT}
For a detailed review see Section 11.13 in
Vol. I of the textbook \\
H. Kleinert,      {\em Gauge Fields in Condensed Matter\/},
     Vol.\ I  {\em Superflow and Vortex Lines\/}, pp. 1--744,
     Vol.\ II  {\em Stresses and Defects\/},
     World Scientific, Singapore 1989, pp. 744-1443
(pubs\#b1).


\bibitem{camlec}
For the wide  use of these generalized $ \delta$-functions
in many fields of physics see the textbook in \cite{seeKT},
and the Cambridge lectures
\\
 H. Kleinert, {\em
 Theory of Fluctuating Nonholonomic Fields and Applications:
    Statistical Mechanics of Vortices and Defects, and
 New Physical
      Laws in Spaces with Curvature and Torsion\/}
	in: Proceedings of a NATO Advanced Study Institute on
	Formation and Interactions of Topological Defects
at the University of Cambridge, England, 1996
(cond-mat/9503030)
\bibitem{cole}
See Section 11.15 in Vol. I of the textbook in \cite{camlec}.
\bibitem{wormh}
See Eq.~(11A.60) in Vol. II of the textbook in \cite{camlec}.

\bibitem{GaugeF}
See Chapter 14 of Vol. II in the textbook in \cite{camlec}.
\bibitem{langh}
H. Kleinert,
     Phys.\ Lett.\  A {\bf  95}, 381 (1983)
(pubs2\#104);\\
W. Janke and H. Kleinert,
     Phys.\ Lett.\  A {\bf 105}, 134 (1984)
(pubs2\#120);
     Phys.\ Lett.\  A {\bf 114}, 255 (1986)
(pubs2\#135).

\bibitem{twostep}
H. Kleinert,
 Phys.\ Lett.\ A {\bf 130}, 443(1988)
(pubs2\#174);
 Phys.\ Lett.\ A {\bf 136}, 468 (1989)
(pubs2\#183);
\\
W. Janke and H. Kleinert,
        Phys.\ Rev.\ Lett.\ {\bf 61}(20), 2344 (1988)
(pubs2\#179);
    Phys.\ Rev.\ A {\bf 41}, 6848 (1990)
(pubs2\#189).

\bibitem{polkl}
 A. Polyakov,
A. M. Polyakov, {\it Nucl. Phys.} B {\bf 268}, 406  (1986);
H. Kleinert,
     Phys.\ Lett.\   {\bf B174}, 335 (1986).
\bibitem{klch}
  H. Kleinert and A. Chervyakov,
Phys. Lett. B {\bf 381}, 286 (1996)
(hep-th/9601030).


\end{thebibliography}
\end{document}